\documentclass[twoside,11pt]{article}

\usepackage{jmlr2e}

\jmlrheading{142}{2021}{1-6}{2nd Workshop on Diversity in Artificial Intelligence (AIDBEI)}{10/00}{Thomas Y. Chen}

\ShortHeadings{PreDefense}{Chen}
\firstpageno{1}

\begin{document}

\title{PreDefense: Defending Underserved AI Students and Researchers from Predatory Conferences}

\author{\name Thomas Y. Chen \email thomaschen7@acm.org \\
       \addr Academy for Mathematics, Science, and Engineering
       Rockaway, NJ 07866, USA}

\editor{Deepti Lamba and William H. Hsu}

\maketitle

\begin{abstract}
Mentorship in the AI community is crucial to maintaining and increasing diversity, especially with respect to fostering the academic growth of underserved students. While the research process itself is important, there is not sufficient emphasis on the submission, presentation, and publication process, which is a cause for concern given the meteoric rise of predatory scientific conferences, which are based on profit only and have little to no peer review. These conferences are a direct threat to integrity in science by promoting work with little to no scientific merit. However, they also threaten diversity in the AI community by marginalizing underrepresented groups away from legitimate conferences due to convenience and targeting mechanisms like e-mail invitations. Due to the importance of conference presentation in AI research, this very specific problem must be addressed through direct mentorship. In this work, we propose PreDefense, a mentorship program that seeks to guide underrepresented students through the scientific conference and workshop process, with an emphasis on choosing legitimate venues that align with the specific work that the students are focused in and preparing students of all backgrounds for future successful, integrous AI research careers.
\end{abstract}

\begin{keywords}
  predatory conferences, artificial intelligence, scientific integrity
\end{keywords}

\section{Introduction}

The existence and continued propagation of predatory scientific conferences and journals is a concerning development for all those that have an interest in scientific integrity. The idea of "predatory" journals and conferences was first introduced to the mainstream by Jeffrey Beall, a librarian. Predatory conferences are "thought to primarily seek profits, in a pay-to-play model where researchers give money to speak at an event" and the organizers harbor "little concern for the quality or rigor of the abstracts they accept or the speakers they invite" \cite{cobey2017conference}. Therefore, the sole purpose of these events, which have troublingly short review periods, is to generate revenue. This has led to amplified criticism of open-access publishing models in general \cite{cortegiani2020predatory}. In 2012, Beall created Beall's List, a registry of journals, conferences, and publishers he deemed to be predatory or likely so; at the time, he was also an associate professor at the University of Colorado Denver \cite{beall2012beall, beall2015predatory}. While the list was not perfect and received a range of criticism, such as that Beall engaged in discrimination against publishers originating in developing countries, it brought an important topic of discussion to the forefront and was praised by many academics as the gold standard for the identification of bogus publishers \cite{kimotho2019storm, somoza2016presence, strielkowski2017predatory}. Other lists have since been developed (e.g. Kscien's List) to tackle this problem \cite{kakamad2019kscien}. It has been documented that even many high-level academics are not fully aware of "predatory science," and that therefore there must be more education and awareness brought to the matter \cite{lang2019approach}.

As these conferences do not incorporate robust peer-review systems, they can promote pseudoscience and are a threat to scientific integrity as a whole; however, specifically in terms of underserved students, they pose a unique threat. Unfortunately, these predatory venues often target early-career researchers and underserved students, who do not know any better than to submit work and are tempted by the opportunity to boost their curricula vitae \cite{mouton2017extent}. For example, the organizers often send e-mail invitations soliciting submissions \cite{asadi2019invitation}. These unsuspecting victims end up paying steep prices for extremely low-quality or even nonexistent conferences. Predatory conferences often target researchers in low-income areas and developing countries, and they have become more adept at designing websites and other material that give them a guise of legitimacy; at first glance, they may not seem predatory \cite{memon2018predatory, mouton2017extent}. In this paper, we focus on predatory conferences specifically and propose a mechanism to combat their success in deceiving well-intentioned students. These conferences largely relate to machine learning and artificial intelligence, as these fields often prioritize presenting at conferences and publishing in their proceedings. The seeming pressure and need to present at "prestigious" venues perpetuates the cycle of students being drawn into predatory conferences. Predatory venues have the consequence of decreasing diversity in the field of AI because those in underrepresented populations attend low-quality conferences and are discouraged from further research. By not being able to present at and participate in mainstream ML conferences such as NeurIPS, ICML, ICLR, etc. and instead continuing to be trapped in and solicited by conferences with practically 100\% acceptance rates, they do not have the opportunity to improve their work and obtain valuable feedback. 

However, it must also be noted that some academics and students are aware of the illegitimacy and possible illegality of particular venues and instead utilize them to bolster their credentials and inflate their CVs: harming both themselves and the entire scientific community by presenting and "publishing" work with no scientific merit \cite{darbyshire2020hitting}. This alternative mindset also hurts underrepresented students because when they consistently submit to conferences with no standards for scientific rigor, they do not have motivation to improve and continue to inflate their CVs while not gaining any valuable insights, perpetuating the cycle. In the absence of immediate legal mechanisms to prosecute these publishers and organizers, it is important that we educate underserved students, particularly, on the most accurate and realistic conferences and workshops to submit their machine learning and artificial intelligence research papers to: both for the benefit of their scientific and professional careers and for the integrity and implications of computer science research. To that end, we propose a mentoring program targeted at underrepresented communities to primarily guide them through the conference submission and publication process. It is focused on late-stage research paper development and the entire submission, review, acceptance, presentation, and publication process. The proposed mentoring program is named PreDefense (with the first "e" being pronounced as a soft vowel). The aim is to maintain and increase a sense of belonging among students from underserved populations while simultaneously limiting research misconduct perpetuated by predatory conferences at-large.\\

\section{Case in Point: WASET}
Before introducing PreDefense, we examine one example of a predatory publisher and distributor of scientific conferences. Namely, we discuss the so-called World Academy of Science, Engineering, and Technology (WASET) \footnote{www.waset.org}, which is a confirmed predatory organizer \cite{cress2017predatory}. On its website, the organization advertises thousands of conferences at various locations across the world. In fact, in each location, there are hundreds of conferences occurring during the same two day period. For example, during January 14th and 15th of 2021 in Singapore, there are "conferences" being held with focuses in topics as disparate as water treatment for hospitals, agronomy, and Islamic architecture. Most notably, there are a wide variety of conferences listed that relate to computer science, machine learning, and AI in general. In Zurich on the same two dates, listed are two dubiously named conferences named "Deep Learning for Pattern Recognition Conference" and "Deep Learning Technologies for Pattern Recognition Conference" — almost identical in name. WASET is only one example of predatory organizing, though it is more high-profile.

\section{PreDefense: How it Will Work}
PreDefense will be offered as a mentorship program that institutions, particularly ones that involve undergraduates, can sign up for. It will start with universities located in low-income and underrepresented areas, as students there will be more likely to not be aware of predatory conferences and why they are problematic. PreDefense will also be particularly directed towards interested institutions in developing countries. Students will be able to subsequently sign up for the program, whenever they are nearing the culmination of the research and paper-writing process. 

The mentorship program would be a part of a nonprofit organization, and underserved students would be able to participate at no cost of their own. Mentors will be educated in and well-versed in not only in the identification of bogus conferences but also the entire conference submission and acceptance process from start to end. There are a number of factors that aid the mentors in identifying fake conferences and recommending appropriate venues. For instance, advertisement emails, virtual conferences without presentation requirements, broad topic areas, changes in dates, mystery regarding committeemembers as well as geographic homogeneity among them, free offers, and dubious payment methods are all signs of a predatory conference \cite{asadi2018fake}. Mentors will be familiar with these signs and be able to navigate the students.

However, the role of the mentor extends far beyond simply recommending conferences. Firstly, the mentors will aid mentees in finalizing their papers. Especially for students in developing countries that have more limited skills in technical writing in English, mentors will have the expertise to guide them in this way. Then, the mentors will gain a solid grasp on the material that the research paper incorporates and recommend any changes. Furthermore, they will begin the process of searching for applicable conferences and workshops. Mentees will be given a wide variety of choices and mentors will explain the benefits and specialties of each. At this time, mentees will be taught specifically how to avoid predatory conferences; this is an important skill they will need to retain after the program in order to succeed academically in the future, as well as to not lend credibility to illegitimate organizers. This is perhaps the most crucial juncture in the mentorship process, due to the ever-increasing pervasiveness of fake conferences.

Subsequently, mentors will guide mentees through the submission procedure, which is useful especially if these are the mentees' first times. The mentor and mentee remain in contact during the review period.

Upon acceptance, mentors will guide mentees through devising a professional presentation. This is also helpful for underserved students that speak English as a second language. Additionally, the conference registration procedure is covered. 

By the conclusion of the session, the primary goal is for mentees to have submitted to a reputable conference, even if their paper was rejected. They are to have learned how to differentiate between predatory conferences and legitimate ones. If they were fortunate to have the opportunity to present, that is yet another skill they carry forward following the closure of the mentorship program.

We expect that this will be an 8-month process at the most, particularly for large conferences such as NeurIPS. In other circumstances, the individual program for the mentee can last as short as 3  months. This is especially true for some workshops, such as the AI for Earth Sciences \footnote{https://ai4earthscience.github.io/neurips-2020-workshop/} workshop at NeurIPS. The interaction between the mentor and mentee will aspirationally be in-person in a post-COVID-19 world, principally due to previous literature highlighting the shortfalls of virtual mentorship, including cultural differences and disparities in internet access \cite{pillon2013mentoring, kafai2013cascading, naggitacode}.

Following preliminary success, the program will expand to universities that may not be traditionally considered "underserved." This remains an important objective because, as previously mentioned, even many high-profile academics do not have sufficient knowledge regarding predatory conferences, given that it is a relatively new development. It has been found that even researchers from top-tier universities like Stanford, Yale, and Harvard submit to predatory journals, due to the organizers' ever-improving mechanisms to seem legitimate \cite{mccrostie2020our}. In any case, the prevalence of predatory organizers is only increasing, and the need for such mentorship programs and targeted education systems specifically on this topic is needed now more than ever \cite{mccrostie2018predatory}. The outsized importance of conferences and workshops in AI research, in comparison to journals alone, motivates this solution.

\section{Conclusion}
The proposed mentorship program, PreDefense, has the principle aim of increasing diversity in AI, which is increasingly under threat from the prevalence of "predatory" conferences. Because there must be mentorship directed towards this issue specifically, PreDefense provides the best proposed service to date. When early-stage researchers, especially students, are encouraged to submit their work to legitimate venues and are educated on the importance of doing so in a healthy manner, the overall state of AI research is promoted. Simultaneously, students from a wide range of backgrounds will be brought into the field by being presented with a clear pipeline for presentation and publication, relieved of the confusion sown by malicious organizers.



\vskip 0.2in
\bibliography{sample}

\begin{thebibliography}{20}
\providecommand{\natexlab}[1]{#1}
\providecommand{\url}[1]{\texttt{#1}}
\expandafter\ifx\csname urlstyle\endcsname\relax
  \providecommand{\doi}[1]{doi: #1}\else
  \providecommand{\doi}{doi: \begingroup \urlstyle{rm}\Url}\fi

\bibitem[Asadi(2019)]{asadi2019invitation}
Amin Asadi.
\newblock Invitation to speak at a conference: The tempting technique adopted
  by predatory conferences’ organizers.
\newblock \emph{Science and engineering ethics}, 25\penalty0 (3):\penalty0
  975--979, 2019.

\bibitem[Asadi et~al.(2018)Asadi, Rahbar, Rezvani, and Asadi]{asadi2018fake}
Amin Asadi, Nader Rahbar, Mohammad~Javad Rezvani, and Fahime Asadi.
\newblock Fake/bogus conferences: Their features and some subtle ways to
  differentiate them from real ones.
\newblock \emph{Science and engineering ethics}, 24\penalty0 (2):\penalty0
  779--784, 2018.

\bibitem[Beall(2012)]{beall2012beall}
Jeffrey Beall.
\newblock Beall’s list of predatory publishers 2013.
\newblock \emph{Scholarly Open Access}, 2012.

\bibitem[Beall(2015)]{beall2015predatory}
Jeffrey Beall.
\newblock Predatory journals and the breakdown of research cultures.
\newblock \emph{Information development}, 31\penalty0 (5):\penalty0 473--476,
  2015.

\bibitem[Cobey et~al.(2017)Cobey, Mazzarello, Stober, Hutton, Moher, and
  Clemons]{cobey2017conference}
Kelly~D Cobey, S~Mazzarello, C~Stober, B~Hutton, D~Moher, and M~Clemons.
\newblock Is this conference for real? navigating presumed predatory conference
  invitations.
\newblock \emph{Journal of Oncology Practice}, 13\penalty0 (7):\penalty0
  410--413, 2017.

\bibitem[Cortegiani et~al.(2020)Cortegiani, Manca, and
  Giarratano]{cortegiani2020predatory}
Andrea Cortegiani, Andrea Manca, and Antonino Giarratano.
\newblock Predatory journals and conferences: why fake counts.
\newblock \emph{Current Opinion in Anesthesiology}, 33\penalty0 (2):\penalty0
  192--197, 2020.

\bibitem[Cress(2017)]{cress2017predatory}
Phaedra~E Cress.
\newblock Are predatory conferences the dark side of the open access movement?,
  2017.

\bibitem[Darbyshire et~al.(2020)Darbyshire, Hayter, Frazer, Ion, and
  Jackson]{darbyshire2020hitting}
Philip Darbyshire, Mark Hayter, Kate Frazer, Robin Ion, and Debra Jackson.
\newblock Hitting rock bottom: The descent from predatory journals and
  conferences to the predatory phd, 2020.

\bibitem[Kafai et~al.(2013)Kafai, Griffin, Burke, Slattery, Fields, Powell,
  Grab, Davidson, and Sun]{kafai2013cascading}
Yasmin Kafai, Jean Griffin, Quinn Burke, Michelle Slattery, Deborah Fields,
  Rita Powell, Michele Grab, Susan Davidson, and Joseph Sun.
\newblock A cascading mentoring pedagogy in a cs service learning course to
  broaden participation and perceptions.
\newblock In \emph{Proceeding of the 44th ACM technical symposium on Computer
  science education}, pages 101--106, 2013.

\bibitem[Kakamad et~al.(2019)Kakamad, Mohammed, Najar, Qadr, Ahmed, Mohammed,
  Salih, Hassan, Mikael, Kakamad, et~al.]{kakamad2019kscien}
Fahmi~H Kakamad, Shvan~H Mohammed, Kayhan~A Najar, Goran~A Qadr, Jaafar~O
  Ahmed, Karukh~K Mohammed, Rawezh~Q Salih, Marwan~N Hassan, Tomas~M Mikael,
  Suhaib~H Kakamad, et~al.
\newblock Kscien's list; a new strategy to hoist predatory journals and
  publishers.
\newblock \emph{International Journal of Surgery Open}, 17:\penalty0 5--7,
  2019.

\bibitem[Kimotho(2019)]{kimotho2019storm}
Stephen~Gichuhi Kimotho.
\newblock The storm around beall’s list: a review of issues raised by
  beall’s critics over his criteria of identifying predatory journals and
  publishers.
\newblock \emph{African Research Review}, 13\penalty0 (2):\penalty0 1--11,
  2019.

\bibitem[Lang et~al.(2019)Lang, Mintz, Krentz, and Gill]{lang2019approach}
Raynell Lang, Marcy Mintz, Hartmut~B Krentz, and M~John Gill.
\newblock An approach to conference selection and evaluation: advice to avoid
  “predatory” conferences.
\newblock \emph{Scientometrics}, 118\penalty0 (2):\penalty0 687--698, 2019.

\bibitem[McCrostie(2018)]{mccrostie2018predatory}
James McCrostie.
\newblock Predatory conferences: a case of academic cannibalism.
\newblock \emph{International Higher Education}, \penalty0 (93):\penalty0 6--8,
  2018.

\bibitem[McCrostie(2020)]{mccrostie2020our}
James McCrostie.
\newblock Our predatory conference problem.
\newblock In \emph{Corruption in Higher Education}, pages 43--48. Brill Sense,
  2020.

\bibitem[Memon and Azim(2018)]{memon2018predatory}
Aamir~Raoof Memon and Muhammad~Ehab Azim.
\newblock Predatory conferences: Addressing researchers from developing
  countries.
\newblock \emph{J Pak Med Assoc}, 68\penalty0 (11):\penalty0 1691--1695, 2018.

\bibitem[Mouton and Valentine(2017)]{mouton2017extent}
Johann Mouton and Astrid Valentine.
\newblock The extent of south african authored articles in predatory journals.
\newblock \emph{South African Journal of Science}, 113\penalty0 (7-8):\penalty0
  1--9, 2017.

\bibitem[Naggita(2020)]{naggitacode}
Keziah Naggita.
\newblock Code for research papers (c4rp).
\newblock 2020.

\bibitem[Pillon and Osmun(2013)]{pillon2013mentoring}
Sylvia Pillon and WE~Osmun.
\newblock Mentoring in a digital age.
\newblock \emph{Canadian Family Physician}, 59\penalty0 (4):\penalty0 442--444,
  2013.

\bibitem[Somoza-Fern{\'a}ndez et~al.(2016)Somoza-Fern{\'a}ndez,
  Rodr{\'\i}guez-Gair{\'\i}n, and Urbano]{somoza2016presence}
Marta Somoza-Fern{\'a}ndez, Josep-Manuel Rodr{\'\i}guez-Gair{\'\i}n, and
  Crist{\'o}bal Urbano.
\newblock Presence of alleged predatory journals in bibliographic databases:
  Analysis of beall's list.
\newblock \emph{El profesional de la informaci{\'o}n}, 25\penalty0 (5), 2016.

\bibitem[Strielkowski(2017)]{strielkowski2017predatory}
Wadim Strielkowski.
\newblock Predatory journals: Beall's list is missed.
\newblock \emph{Nature}, 544\penalty0 (7651):\penalty0 416--416, 2017.

\end{thebibliography}

\end{document}